\newif\ifsingle

\ifsingle
\documentclass[11pt,draftclsnofoot, onecolumn]{IEEEtran}		
\else		
\documentclass[10pt,final, twocolumn]{IEEEtran}
\fi

\usepackage{times}
\usepackage{amsmath,dsfont}
\usepackage{amssymb,amsthm}
\usepackage{epsfig,verbatim}
\usepackage{subfigure}
\usepackage{setspace}
\usepackage{color}
\usepackage{cite}
\usepackage{epstopdf}
\usepackage{graphics}
\usepackage{accents}
\usepackage{acronym}
\usepackage[bookmarks,colorlinks]{hyperref}
\usepackage{booktabs}
\usepackage{mathtools}
\usepackage{algorithm}
\usepackage{algorithmic}
\usepackage{enumitem}
\usepackage{pstool}
\usepackage[linewidth=0.5pt]{mdframed}

\ifsingle

\else

\fi 

\acrodef{adc}[ADC]{Analog-to-Digital Convertor}
\acrodef{dac}[DAC]{digital-to-analog convertor}
\acrodef{cs}[CS]{Compressed Sensing}
\acrodef{dtft}[DTFT]{discrete-time Fourier transform}
\acrodef{dnn}[DNN]{deep neural network} 
\acrodef{csi}[CSI]{channel state information}
\acrodef{map}[MAP]{maximum a-posteriori probability}
\acrodef{snr}[SNR]{signal-to-noise ratio}
\acrodef{sinr}[SINR]{signal-to-interference-and-noise ratio}
\acrodef{bs}[BS]{Base Station} 
\acrodef{em}[EM]{electromagnetic} 
\acrodef{iot}[IOT]{Interent of Things}
\acrodef{mimo}[MIMO]{multiple-input multiple-output}
\acrodef{mse}[MSE]{mean-squared error}
\acrodef{pdf}[PDF]{probability density function}
\acrodef{rv}[RV]{random variable}
\acrodef{fec}[FEC]{forward error correction}
\acrodef{rs}[RS]{Reed-Solomon}
\acrodef{lti}[LTI]{linear time-invariant}
\acrodef{wss}[WSS]{wide-sense stationary}
\acrodef{psd}[PSD]{power spectral density}
\acrodef{ser}[SER]{symbol error rate} 
\acrodef{ber}[BER]{bit error rate} 
\acrodef{isi}[ISI]{intersymbol interference}  
\acrodef{awgn}[AWGN]{additive white Gaussian noise} 
\acrodef{ut}[UTs]{User Terminals} 
\acrodef{mmw}[mmWave]{millimeter wave}
\acrodef{ris}[RIS]{reconfigurable intelligent surface} 
\acrodef{dma}[DMA]{Dynamic Metasurface Antenna} 
\acrodef{5G}{fifth generation}
\acrodef{pa}[PA]{power amplifier}

\IEEEoverridecommandlockouts

\title{Extremely Large-Scale Dynamic Metasurface Antennas for 6G Near-Field Networks: Opportunities and Challenges}

\author{  
	\IEEEauthorblockN{Haiyang Zhang, Nir Shlezinger, Giulia Torcolacci, Francesco Guidi, Anna Guerra,  Qianyu Yang, \\ Mohammadreza F. Imani,  Davide Dardari, and Yonina C. Eldar\\
	} 

}
 
\begin{document}
	
	\maketitle
 	\pagestyle{plain}  
\thispagestyle{plain} 


\begin{abstract}

6G networks will need to support higher data rates, high-precision localization, and imaging capabilities. Near-field technologies, enabled by extremely large-scale (XL)-arrays, are expected to be essential physical-layer solutions to meet these ambitious requirements. However, implementing XL-array systems using traditional fully-digital or hybrid analog/digital architectures poses significant challenges due to high power consumption and implementation costs. Emerging XL-dynamic metasurface antennas (XL-DMAs) 
provide a promising alternative, enabling ultra-low power and cost-efficient solutions, 
making them ideal candidates for 6G near-field networks. In this article, we discuss the opportunities and challenges of XL-DMAs employed in 6G near-field networks. We first outline the fundamental principles of XL-DMAs and present the specifics of the near-field model of XL-DMAs. 
We then highlight several promising applications that might benefit from XL-DMAs, including near-field communication, localization, and imaging. Finally, we discuss several open problems and potential future directions that should be addressed to fully exploit the capabilities of XL-DMAs in the next 6G near-field networks.

\end{abstract}

\bstctlcite{IEEEexample:BSTcontrol}

\section{Introduction}

Next generations of wireless networks are expected to support peak data rates exceeding 100 Gigabits per second, high-precision localization within 10 cm, and advanced imaging capabilities to enable future applications such as autonomous vehicles, smart cities, and industrial automation \cite{10536135}. To meet these ambitious requirements, the use of upper mid-band, millimeter-wave, and terahertz frequencies, alongside extremely large-scale (XL) antenna arrays, is expected to play a key role~\cite{bjornson2024towards}. 
The evolution of XL-arrays systems operating in these high-frequency bands for 6G networks naturally transitions the operating regime from far-field to near-field propagation, thanks to the extended Fraunhofer distance, which delineates the boundary between the propagation regions~\cite{10566015}.

In the near-field region, wavefronts can not be approximated as planar, 
unlocking new opportunities for precise beamforming and spatial control~\cite{liu2023near}. This paradigm shift paves the way for innovative advancements in 6G networks, particularly in communication, localization, and imaging applications.
For example, in multi-user communication scenarios, near-field focused beams ensure reliable communication, even when users are located at different distances but share the same angular direction~\cite{zhang2022beam}, and facilitate the integration of communications and sensing~\cite{cong2024near}. For the latter application, exploiting the near-field spherical wavefronts facilitates curvature of arrival (COA)-based localization, which not only enhances localization accuracy but also alleviates the stringent requirements for precise synchronization and/or multiple anchor nodes typically needed in conventional far-field localization techniques~\cite{guerra2021near}. 

Realizing the numerous advantages of XL-arrays for 6G networks necessitates dedicated antenna architectures~\cite{10858129}. The implementation of XL-arrays systems using traditional designs based fully-digital or hybrid analog/digital architectures presents significant challenges due to high power consumption and implementation costs. To face these challenges, dynamic metasurface antennas (DMAs), which is a specific class of holographic metasurfaces, represent an emerging and appealing technology\cite{jabbar202460, Sleasman-2016JAWPL}. Notably, DMAs leverage reconfigurable metamaterial elements to enable flexible wavefront manipulation, adaptive beam steering, and analog-domain signal processing~\cite{williams2022electromagnetic}, while being a scalable, low-cost, and energy-efficient alternative to conventional massive antenna arrays~\cite{shlezinger2021dynamic}. With sub-wavelength element spacing, DMAs can incorporate a larger number of elements within a limited antenna aperture, thereby allowing for greater control over electromagnetic waves. These advantages make DMAs particularly suitable for implementing XL-arrays, namely XL-DMAs, and are thus expected to play a key role for near-field signaling in 6G wireless networks.

\begin{figure*}
	\centering
	\includegraphics[width=0.75\linewidth]{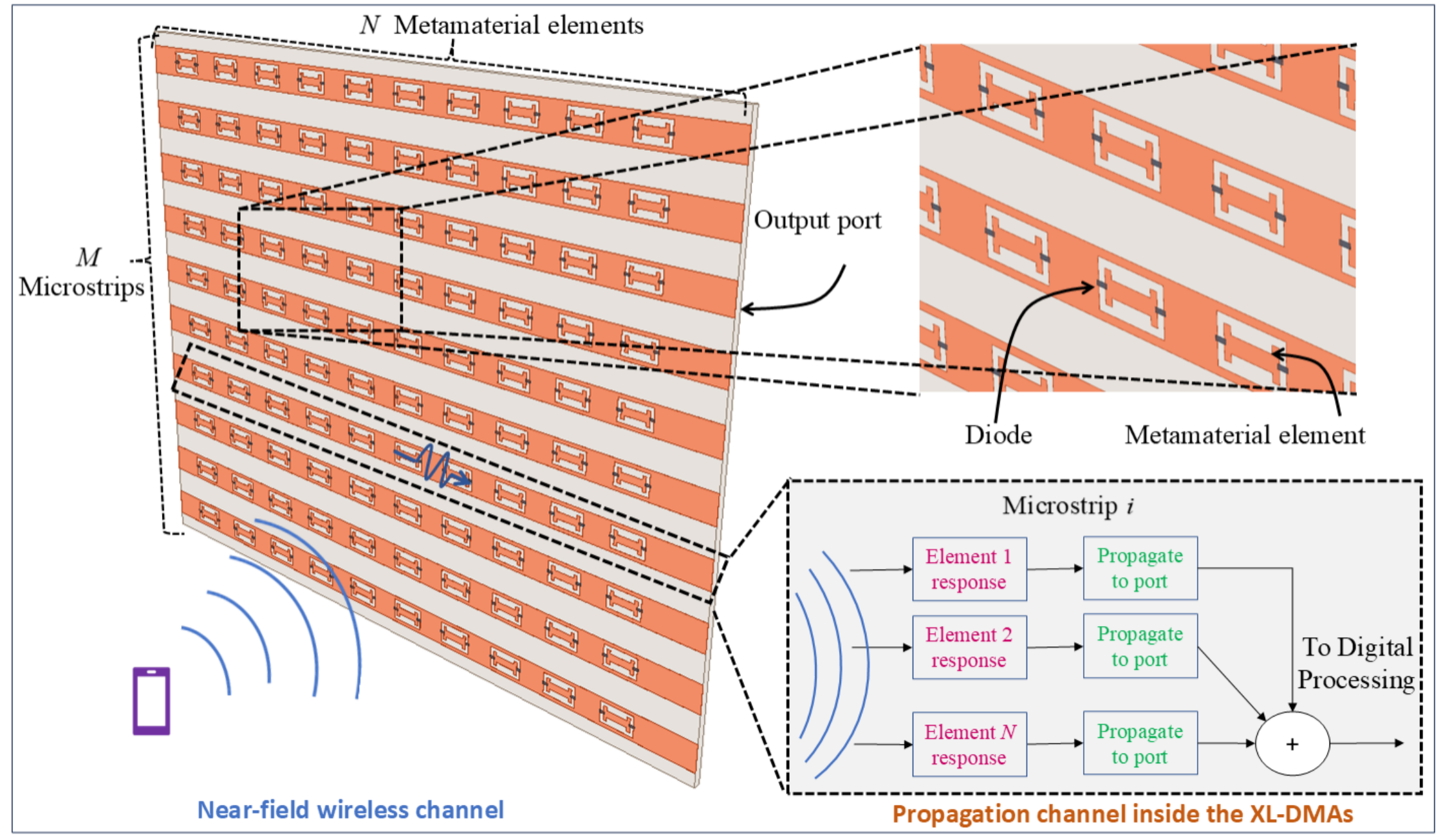}
	\caption{XL-DMAs schematic illustration.}
	\label{fig:DMA_illustration}
\end{figure*}

In this article, we provide an overview of the opportunities and challenges associated with the use of XL-DMAs in 6G near-field networks. We first present the fundamental principles of XL-DMAs and how their structure differs from that of traditional antenna arrays. Next, we incorporate the unique properties of XL-DMAs into wireless channel models and discuss the equivalent near-field channel model. Then, we explore several promising 6G  applications enhanced by XL-DMAs, including $(i)$ Near-field communications with XL-DMAs, emphasizing the advantages of the subwavelength feature of XL-DMAs, which enhances communication capacity in near-field communication scenarios;
$(ii)$ Near-field localization with XL-DMAs, demonstrating the advantages of curvature-of-arrival-based localization techniques; 
and $(iii)$ Near-field imaging with XL-DMAs, leveraging near-field high-rank properties of wireless channels to achieve superior resolution.

We proceed by examining representative case studies for the aforementioned application areas, presenting numerical results to demonstrate the effectiveness of XL-DMAs in enhancing communication, localization accuracy, and imaging quality. We conclude by highlighting key design challenges and research opportunities for XL-DMAs-assisted near-field technologies, including artificial intelligence (AI)-based XL-DMAs configuration, distributed XL-DMAs-empowered near-field operating, polarization diversity with XL-DMAs, as well as hardware implementation considerations. Through these discussions, we establish XL-DMAs as a promising technology for realizing the vision of 6G networks that seamlessly support not only high-data-rate communication but also high-accuracy localization and imaging in near-field environments.



\section{Fundamentals of XL-DMAs }

To present the basics of XL-DMAs, we begin with a brief description of XL-DMAs hardware architectures. Then, we elaborate on the properties of their near-field channel model, setting the groundwork for their application in near-field communication, localization, and imaging.

\subsection{XL-DMA Hardware Architecture}
\label{subsec:DMA_hardware}

DMAs are a novel antenna architecture based on tunable metamaterials, designed to implement extremely large-scale antenna arrays with low power consumption and cost. A typical DMA array consists of multiple independent one-dimensional waveguides~\cite{shlezinger2021dynamic}, often realized using microstrips, each containing numerous sub-wavelength-sized metamaterial elements, as shown in Fig.~\ref{fig:DMA_illustration}. These waveguides are connected to dedicated RF chains for both signal transmission and reception. 
By increasing the number of microstrips or metamaterial units, an XL-DMAs can be formed.

XL-DMAs offer two key advantages compared to conventional antenna architectures considered for XL-arrays: first, they significantly reduce power consumption compared to traditional hybrid architectures by eliminating the need for additional components such as phase shifters and analog combiners, as the metamaterial elements' inherent resonant properties can be leveraged to implement desired beamforming and focusing for transmission and reception. Secondly, the sub-wavelength spacing of metamaterial elements allows more antenna elements to be integrated within the same aperture, which can be leveraged to realize holographic beamforming and focusing, especially when the mutual coupling is accounted for~\cite{11095395}. 



\subsection{Near-Field Channel Model with XL-DMAs}
\label{subsec:DMAIntroc}

	\begin{figure*}
		\centering
		\includegraphics[width= 1\linewidth]{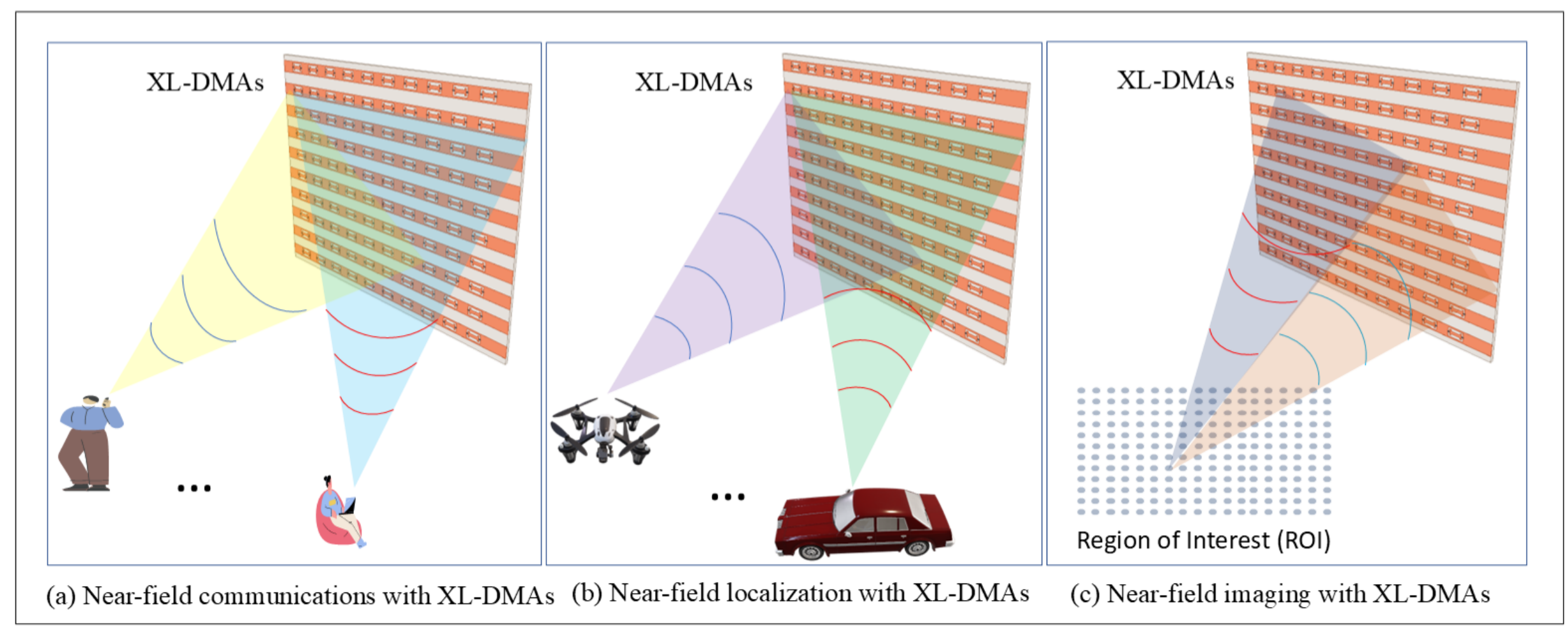}
		\caption{ Typical applications of XL-DMAs in emerging 6G near-field networks. (a) Near-field communications with XL-DMAs; (b) Near-field localization with XL-DMAs;  (c) Near-field imaging with XL-DMAs.}
		\label{fig:application}
	\end{figure*}

As depicted in Fig.~\ref{fig:DMA_illustration}, the near-field channel model of XL-DMAs consists of two components: the wireless channel between the users and the XL-DMA elements, and the propagation channel inside the XL-DMAs. The wireless channel component accounts for the near-field effects arising from the large aperture of XL-DMAs, where electromagnetic waves propagate as spherical wavefronts.
Consequently, the wireless channel component depends on the distance and geometry of the antenna arrays and can be described by a spherical wave propagation model~\cite{zhang2022beam}. Furthermore, due to the large aperture of XL-DMAs, different parts of the antenna array may experience varying channel conditions, introducing spatial non-stationarity effects, which should also be considered when modeling near-field wireless channels.

The propagation channel within XL-DMAs captures both the frequency response of the metamaterial elements and the propagation characteristics inside the microstrips. The frequency response of the metamaterial elements is often modeled based on coupled-dipole formalism as a bandpass filter with tunable Lorentzian resonance responses, enabling control of amplitude and phase~\cite{DSmith-2017PRA}.
The propagation characteristics inside the microstrip depend on the position of each metamaterial element along the microstrip. For example, on the receive side of XL-DMAs, the output of each microstrip is obtained as a linear combination of the signals observed by the metamaterial elements, effectively capturing the propagation characteristics within the microstrip, as shown in Fig.~\ref{fig:DMA_illustration}.

Based on the discussion above, the channel models of XL-DMAs differ from conventional near-field channel models used in fully digital or phase-shifter-based hybrid antenna architectures. For accurate modeling, the unique characteristics of XL-DMAs, such as spherical wave propagation and the inherent propagation channel within the XL-DMA itself, must be taken into account.





\section{Near-Field Applications with XL-DMAs}
\label{sec:appliations}

As highlighted above,  
XL-DMAs represent a promising antenna technology with the potential to meet 6G's ambitious requirements. This section discusses some of their promising applications,
including near-field communication, near-field localization, and near-field imaging.

\subsection{Near-Field Communication}
\label{subsec:DMA-NF}

The spherical wave propagation characteristics offer significant advantages for XL-arrays or XL-MIMO near-field communications, as shown in Fig.~\ref{fig:application}(a). First, due to the phase differences of signals received at various antenna positions, the near-field line-of-sight MIMO channel achieves a higher rank, which boosts communication capacity compared to the far-field scenario. 
Moreover, near-field spherical wave propagation introduces an additional degree of freedom via the distance dimension, alongside angular information. 
This enables beam focusing that allows to precisely focus the signal energy towards designated spatial locations, providing superior control over signal propagation in both the angular and distance dimensions.
 Consequently, beam focusing not only enhances the received signal strength for target users but also effectively mitigates multi-user interference in the joint angle-distance domain, thereby improving the communication system access efficiency~\cite{zhang2022beam,liu2023near}.

The above potential gains, which hold for any XL-MIMO system operating in the near-field region, are naturally supported and facilitated by XL-DMAs. 
First, XL-DMAs allow for scalable and affordable deployments of XL-MIMO arrays. Their support of subwavelength-spaced metamaterial elements facilitates packing a large number of radiating elements within a given aperture, which can be exploited  to increase the channel rank and boosts channel capacity. 
Second, the tunable ability of XL-DMAs elements allows for flexible beam pattern generation, which enhances beam focusing and multi-user interference control, further optimizing performance in near-field XL-MIMO systems.

\subsection{Near-Field Localization}
\label{subsec:DMA-NF}

In 6G networks, as shown in Fig.~\ref{fig:application}(b),  centimeter-level localization is required to support emerging applications such as autonomous driving.
Achieving such high precision, however, is challenging with conventional far-field localization methods.
For instance, angle of arrival (AoA) and time of arrival techniques typically require precise synchronization and multiple access points to provide accurate location estimates, yet maintaining synchronization remains a persistent challenge.

In contrast, near-field localization can utilize CoA methods~\cite{guerra2021near}, which capitalizes on phase coherence between multiple signal arrivals to accurately estimate distance and angle, offering higher precision while reducing the reliance on synchronization. As this operation requires some flexibility in processing the received signals, the use of XL-DMAs, with their tunable metamaterial elements, enhances near-field localization. Specifically, the tunable nature of XL-DMAs allows for more flexible beam shaping in the near-field, enhancing the system's capability to focus and track signals across both angular and distance dimensions. In addition, the beam focusing ability of XL-DMAs reduces multi-user interference  and enhances the signal-to-noise ratio, leading to higher localization accuracy. Moreover, similarly to communications, XL-DMAs' sub-wavelength elements enable denser packing of antenna elements, improving signal resolution and thereby increasing localization accuracy.

\subsection{Near-Field Imaging}
\label{subsec:DMA-NF}

While the benefits of operating in the near-field propagation regime have been studied for communication and localization, the potential advantages of performing holographic imaging within wireless communication networks remain largely unexplored \cite{10566015}. 
Imaging can be regarded as a sensing task, and achieving high-quality imaging is a key objective for future 6G networks, as illustrated in Fig.~\ref{fig:application}(c). Specifically, holographic imaging involves illuminating a region of interest (ROI) with radio frequency or microwave waves, capturing the reflected and scattered signals to estimate the ROI's radar cross section or its scattering coefficients, and using interference patterns along with advanced signal processing to reconstruct a detailed 2D/3D image of the object’s shape and orientation.

Near-field imaging provides enhanced spatial resolution, greater sensitivity, and superior imaging performance.
This is because the near-field spherical wavefront enables more intricate interactions with the ROI, allowing for finer spatial sampling and achieving higher resolution  \cite{10566015}. In contrast, far-field imaging relies on planar wavefront, where resolution is constrained by the electromagnetic waves' divergence. Additionally, near-field imaging benefits from a higher channel rank, as the increased number of distinct signal components improves image reconstruction accuracy and overall quality. The deployment of XL-DMAs further amplifies these advantages \cite{imani2020review}. The tunable elements of XL-DMAs facilitate flexible beam pattern generation, improving beam focusing and multi-user interference control, ultimately leading to a higher-quality image reconstruction. Moreover, the natural support of XL-DMAs in employing a large number of antenna elements enhances the channel rank and allows for an optimized illumination of the ROI. Despite these promising potentials, the full exploitation of near-field imaging with XL-DMAs remains under-explored.





\section{Case Studies}
\label{ssec:Sim}
To provide empirical evidence on the advantages of XL-DMAs-enabled 6G near-field systems, we report numerical results corresponding to representative case studies in near-field communications, localization, and imaging.

    \begin{figure}
	\centering
	\includegraphics[width= 3.8in]{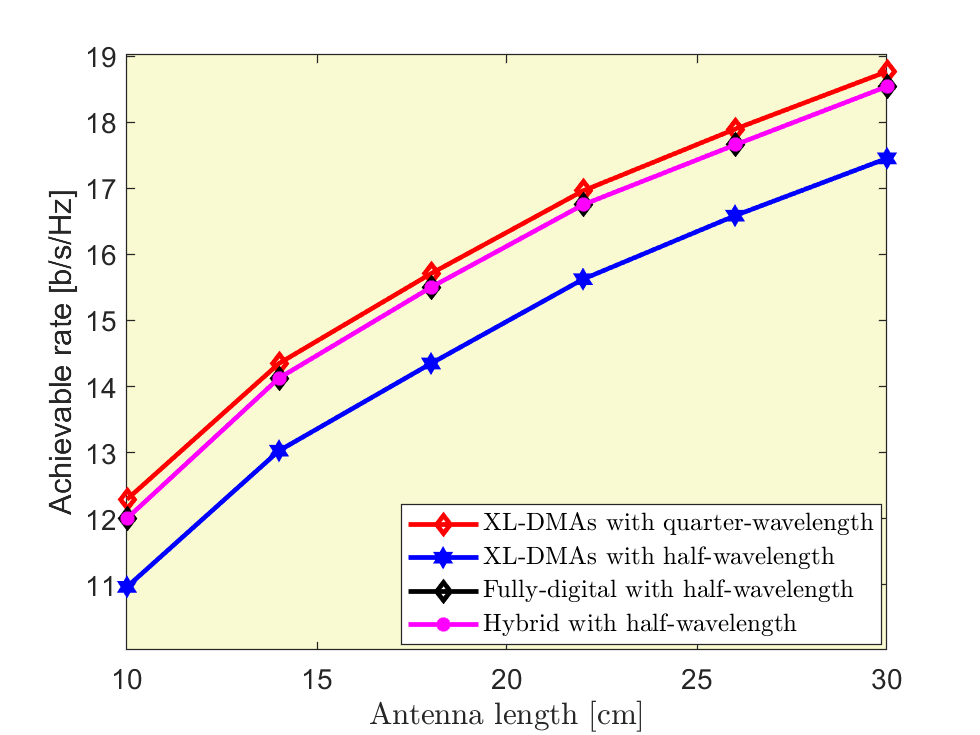}
	\caption{Achievable rate comparison of different antenna architectures with varying antenna lengths. The XL-DMAs are positioned in the $xy$-plane, while the near-field user is located along the $z$-axis at a distance of 150 wavelengths.}
	\label{fig:com}
\end{figure}

	\begin{figure}
		\centering
		\includegraphics[width= 3.8in]{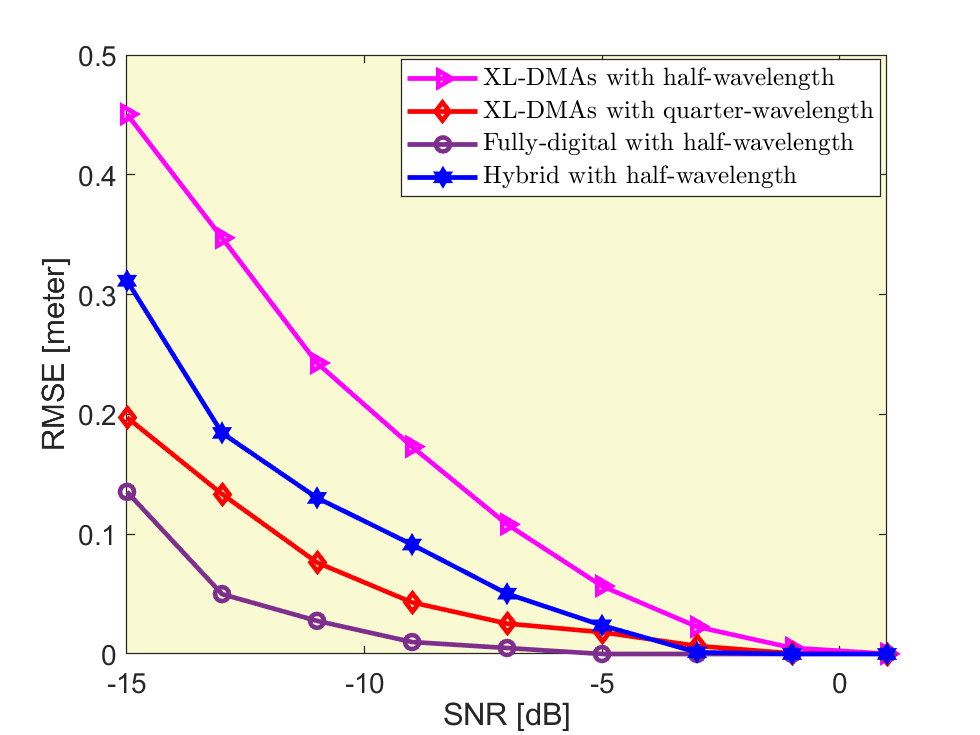}
		\caption{Localization RMSE comparison of different antenna architectures with varying SNR. The antennas, with a length of 35  centimeters, lie in the $yz$-plane. The two users are positioned in the $xy$-plane with the following coordinates: User 1 and User 2 are located at $(0.25~\text{Rayleigh distance}, \pi/6)$ and $(0.25~\text{Rayleigh distance}, \pi/4)$ in polar coordinates, respectively. 
 }
		\label{fig:localization}
	\end{figure}

\begin{figure*}
	\centering
	\includegraphics[width=0.8\linewidth]{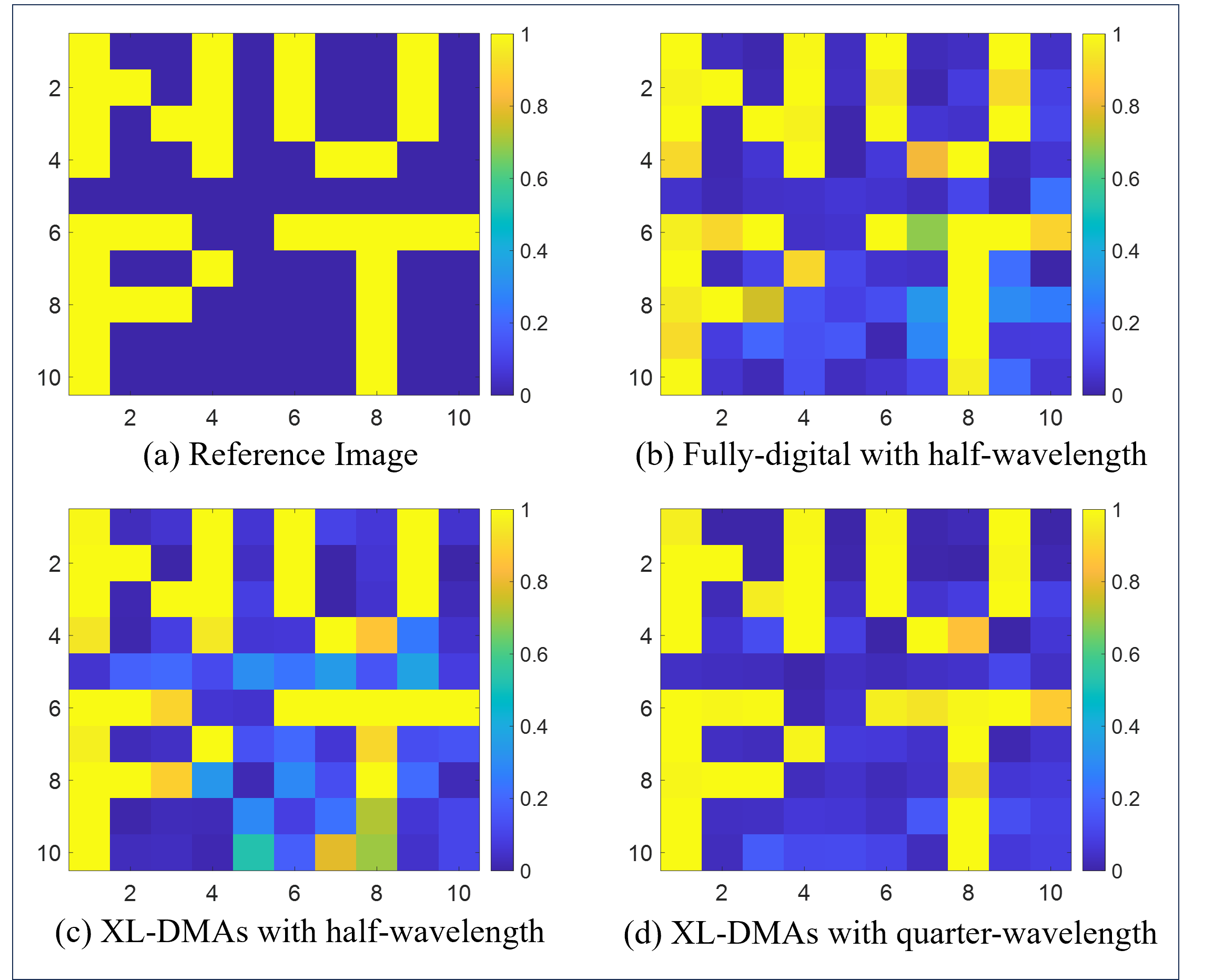}
	\caption{Estimated images under different antenna architectures. (a) Reference image; (b) Estimated image using a fully-digital antenna with half-wavelength spacing; (c) Estimated image using XL-DMAs with half-wavelength spacing; (d) Estimated image using XL-DMAs with quarter-wavelength spacing.}
	\label{fig:image}
\end{figure*}

To demonstrate the potential of XL-DMAs for near-field communications, Fig.~\ref{fig:com} compares the single-user downlink achievable rate performance of XL-DMAs with that of fully-digital and phase shifter-based hybrid antennas, under varying antenna lengths. The simulations are conducted at a carrier frequency of 28 GHz. For fully-digital and hybrid antennas, the antenna element spacing is set to half the wavelength, while for XL-DMAs, the element spacings are configured as one-quarter and half the wavelength, respectively. The XL-DMAs waveguide parameters are consistent with those used in \cite{zhang2022beam},  with the transmit power and noise power set to -13 dBm and -174 dBm, respectively. From Fig.~\ref{fig:com}, it is verified that XL-DMAs with an element spacing of one-quarter of the wavelength outperform both fully-digital and hybrid antennas. This is because the smaller element spacing allows to pack more radiating elements within the same aperture, enhancing performance and confirming the advantages of XL-DMAs in achieving higher communication rate.
Additionally, Fig.~\ref{fig:com} shows that as the antenna length increases, the achievable rate improves significantly across all three antenna architectures. This improvement is attributed to larger antenna apertures, which extend the near-field region and enhance signal strength at the focusing point. 

Fig.~\ref{fig:localization} compares the near-field localization accuracy, measured by the Root Mean Square Error (RMSE), of XL-DMAs, fully-digital arrays, and phase-shifter-based hybrid arrays, under varying Signal-to-Noise Ratio (SNR) in a near-field localization scenario. From Fig.~\ref{fig:localization}, it is evident that XL-DMAs with quarter-wavelength spacing achieve superior localization accuracy compared to hybrid arrays and closely approach the performance of fully-digital arrays. This improvement is attributed to the higher element density enabled by quarter-wavelength spacing and the tunable properties of metamaterials, which enhance spatial resolution and improve the system’s ability to estimate user positions with greater precision. It is also worth noting that, compared to fully-digital arrays, which require the same number of RF chains as antenna elements, XL-DMAs achieve comparable accuracy while requiring significantly fewer RF chains. These results verify the potential of XL-DMAs as an efficient and cost-effective solution for near-field localization applications.

Fig.~\ref{fig:image} illustrates the application of XL-DMAs to near-field imaging, comparing the estimated images generated using fully-digital arrays. The reference image serves as the ground truth for evaluating the imaging accuracy of different antenna architectures. From Fig.~\ref{fig:image}, it can be observed that the estimated image produced using XL-DMAs with half-wavelength spacing is close to the fully-digital image. However, the estimated image produced using XL-DMAs with quarter-wavelength spacing is 
much better than the fully-digital image and resembles the reference image more closely. This implies that XL-DMAs can achieve higher imaging accuracy than fully-digital arrays while requiring significantly fewer RF chains, demonstrating the applicability of XL-DMAs to near-field imaging.

\section{Design challenges and research directions}
\label{sec:directions}

XL-DMAs offer significant benefits to 6G near-field networks, as highlighted above. However, this remains a relatively new area of research, and several open problems and potential future directions need to be explored to fully unlock the potential of XL-DMAs for 6G near-field networks.


 \subsection{AI-Based XL-DMAs Configuration}
\label{ssec:MIMO}

In XL-DMAs-assisted 6G near-field networks, the frequency response of metamaterial elements is integrated into the equivalent near-field channels. As a result, the performance of these networks heavily depends on the configuration of the XL-DMAs. Since each metamaterial element must satisfy the Lorentzian constraint, configuring the parameters of the XL-DMAs becomes particularly challenging, especially for wideband applications. Currently, iterative and mapping optimization methods are used to configure XL-DMA parameters, but these approaches are complex and suboptimal. Therefore, exploiting the capabilities of  AI to intelligently configure the parameters of XL-DMAs for various application tasks represents a promising direction for future 6G research.



\subsection{Distributed XL-DMAs-empowered Near-Field Operating}

Rather than relying on a centralized deployment of XL-DMAs, a distributed XL-DMAs architecture, which involves deploying multiple XL-DMAs across a spatially distributed framework, has the potential to significantly improve coverage in the near-field region and enhance the performance of various near-field applications, such as communication, localization, and imaging. However, this approach introduces several challenges, including the need for precise synchronization and coordination between distributed nodes, efficient channel estimation across spatially separated XL-DMAs, and effective interference management to mitigate signal overlap. Overcoming these challenges and fully harnessing the potential of distributed XL-DMAs in 6G near-field communication networks represents a promising area for future research.

\subsection{Polarization Diversity with XL-DMAs}

XL-DMAs possess the unique capability to generate and manipulate different polarization states of electromagnetic waves with high flexibility. Exploiting this property can significantly enhance the performance of near-field communication, localization, and imaging systems. For instance, polarized beams can reduce interference and improve channel capacity by leveraging orthogonal polarization states to support multiple data streams. Additionally, polarization diversity in near-field imaging and localization can provide valuable polarization-dependent information about objects or environments, enhancing resolution and accuracy. Therefore, future research should investigate strategies to fully leverage the polarization capabilities of XL-DMAs to advance 6G near-field applications.

 \subsection{Hardware Implementation}
\label{ssec:hardware}
 
Hardware developments and over-the-air experiments are critically important in verifying the effectiveness of XL-DMA-assisted near-field communication networks. For example, the high-density integration of metamaterial elements poses significant challenges, as mutual coupling between adjacent elements can degrade performance and introduce design complexities. Additionally, achieving dynamic tunability of metamaterial elements requires flexible control mechanisms that allow for precise and independent amplitude and phase adjustment. Furthermore, the efficiency of power amplifiers is a major concern, as it is typically frequency-dependent and tends to decrease at higher carrier frequencies. All these challenges must be carefully addressed to ensure the successful implementation of XL-DMAs in 6G near-field communication networks.


\section{Conclusion}

Near-field propagation, facilitated by XL-arrays, offers significant advantages for 6G networks, enabling higher data rates, precise localization, and advanced imaging capabilities that surpass those of 5G. However, implementing XL-array systems using traditional fully-digital or hybrid analog/digital architectures is challenging due to high power consumption and implementation costs. Alternatively, XL-DMAs can realize XL-arrays with ultra-low power consumption and reduced costs, making them ideal candidates for 6G near-field networks. In this article, we provided an overview of the opportunities and challenges of XL-DMA-powered 6G near-field networks. We first introduced the fundamental principles of XL-DMAs and presented the unique near-field channel model of XL-DMAs, which captures the near-field spherical wavefront feature and signal propagation characteristics within the microstrip. We then highlighted several promising application scenarios for XL-DMAs, including near-field communication, localization, and imaging. Finally, we concluded with several open problems and potential future directions, which are expected to pave the way for implementing XL-DMA-assisted 6G near-field networks.

	\bibliographystyle{IEEEtran}
	\bibliography{IEEEabrv,refs}


\begin{IEEEbiographynophoto}{Haiyang Zhang}(haiyang.zhang@njupt.edu.cn) 
 is an Assistant Professor in the School of Communication and Information Engineering, Nanjing University of Posts and Telecommunications, China.
\end{IEEEbiographynophoto}	
\vskip -2\baselineskip plus -1fil

\begin{IEEEbiographynophoto}{Nir Shlezinger}(nirshl@bgu.ac.il)   is an Assistant Professor in the School of Electrical and Computer Engineering in Ben-Gurion University, Israel. 
\end{IEEEbiographynophoto}
\vskip -2\baselineskip plus -1fil

 \begin{IEEEbiographynophoto}{Giulia Torcolacci}
 (g.torcolacci@unibo.it) is a Ph.D. candidate at the University of Bologna, Italy, within the Department of Electrical, Electronic, and Information Engineering “Guglielmo Marconi”, and an affiliate at CNIT-WiLab, Bologna, Italy. 
 \end{IEEEbiographynophoto}
 \vskip -2\baselineskip plus -1fill

\begin{IEEEbiographynophoto}{Francesco Guidi}(francesco.guidi@cnr.it)
received his Ph.D. degree in electronics, telecommunications, and information
technologies from Ecole Polytechnique Paris, France and from the University of Bologna,
Italy. He is currently a Senior Researcher with IEIIT-CNR, Italy.
\end{IEEEbiographynophoto}
\vskip -2\baselineskip plus -1fill

\begin{IEEEbiographynophoto}{Anna Guerra} (anna.guerra3@unibo.it) is an Associate Professor in the Department of Electrical, Electronic, and Information Engineering "Guglielmo Marconi" (DEI) at the University of Bologna, Italy. 
\end{IEEEbiographynophoto}
\vskip -2\baselineskip plus -1fil

 \begin{IEEEbiographynophoto}{Qianyu Yang}(2020010207@njupt.edu.cn)
 is a 
 Ph.D. candidate in the School of Communication and Information Engineering, Nanjing University of Posts and Telecommunications, China.  
 \end{IEEEbiographynophoto}
 \vskip -2\baselineskip plus -1fill

 \begin{IEEEbiographynophoto}{Mohammadreza F. Imani}(mohammadreza.imani@asu.edu)
 is an Assistant Professor in the ECEE school, Arizona State University, USA.  
 \end{IEEEbiographynophoto}
 \vskip -2\baselineskip plus -1fill
	
\begin{IEEEbiographynophoto}{Davide Dardari} (davide.dardari@unibo.it) is a full professor in the Department of Electrical, Electronic, and Information Engineering "Guglielmo Marconi" (DEI) at the University of Bologna, and affiliate at WiLAB-CNIT, Italy.  He is an IEEE Fellow.
\end{IEEEbiographynophoto}	
\vskip -2\baselineskip plus -1fill

\begin{IEEEbiographynophoto}{Yonina C. Eldar} (yonina.eldar@weizmann.ac.il)
is a Professor in the Department of Math and Computer Science, Weizmann Institute of Science, Israel, where she heads the center for Biomedical Engineering and Signal Processing. She is a member of the Israel Academy of Sciences and Humanities, an IEEE Fellow and a EURASIP Fellow.
\end{IEEEbiographynophoto}	
	
\end{document}